\documentclass{article}

\usepackage{arxiv}

\usepackage[utf8]{inputenc} 
\usepackage[T1]{fontenc}    
\usepackage{hyperref}       
\usepackage{url}            
\usepackage{booktabs}       
\usepackage{amsfonts}       
\usepackage{nicefrac}       
\usepackage{microtype}      
\usepackage{lipsum}		
\usepackage{graphicx}
\usepackage{natbib}
\usepackage{doi}
\usepackage{xcolor}        
\usepackage{amsmath}
\usepackage{booktabs}
\usepackage{amssymb}
\usepackage[ruled,vlined]{algorithm2e} 
\usepackage{float}

\title{From Learning to Control: Data-Driven Multi-Agent Reinforcement Learning for Multivariable Control in a Microalgae Bioprocess}

\author{ \href{https://orcid.org/0000-0003-1484-6923}{\includegraphics[scale=0.06]{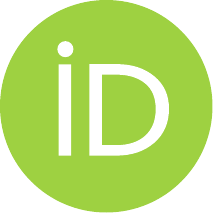}\hspace{1mm}Juan D. Gil} \\
	Department of Informatics \\
    University of Almería, CIESOL, ceiA3, \\
    Almería, 04120, Spain \\
	\texttt{juandiego.gil@ual.es} \\
	\And
	\href{https://orcid.org/0000-0003-0274-2852}{\includegraphics[scale=0.06]{orcid.pdf}\hspace{1mm}Ehecatl Antonio Del Rio Chanona} \\
	Sargent Centre for Process Systems Engineering\\
	Imperial College London\\
	SW7 2AZ, London, UK\\
	\texttt{a.del-rio-chanona@imperial.ac.uk} \\
	  \AND
      \href{https://orcid.org/0000-0001-5312-0776}{\includegraphics[scale=0.06]{orcid.pdf}\hspace{1mm}José Luis Guzmán} \\
	Department of Informatics \\
    University of Almería, CIESOL, ceiA3, \\
    Almería, 04120, Spain \\
	\texttt{joseluis.guzman@ual.es} \\
    	  \And
      \href{https://orcid.org/0000-0002-3349-7506}{\includegraphics[scale=0.06]{orcid.pdf}\hspace{1mm}Manuel Berenguel} \\
	Department of Informatics \\
    University of Almería, CIESOL, ceiA3, \\
    Almería, 04120, Spain \\
	\texttt{beren@ual.es} \\
}

\renewcommand{\shorttitle}{\textit{arXiv} Template}

\hypersetup{
pdftitle={From Learning to Control: Data-Driven Multi-Agent Reinforcement Learning for Multivariable Control in a Microalgae Bioprocess},
pdfsubject={q-bio.NC, q-bio.QM},
pdfauthor={David S.~Hippocampus, Elias D.~Striatum},
pdfkeywords={First keyword, Second keyword, More},
}

\begin{document}
\maketitle

\begin{abstract}
	Effective control of bioprocesses is particularly challenging due to the intrinsic nonlinearity and dynamic variability of living-cell systems. In microalgae-based photobioreactors (PBRs), maintaining stable $\mathrm{pH}$ and dissolved oxygen ($\mathrm{DO}$) levels is critical for optimal growth and productivity, yet their strong coupling and sensitivity to environmental fluctuations make multivariable control difficult. This study proposes a novel hybrid offline-online Multi-Agent Reinforcement Learning (MARL) framework for simultaneous $\mathrm{pH}$ and $\mathrm{DO}$ regulation, leveraging Deep Deterministic Policy Gradient (DDPG) agents to achieve a fully data-driven and model-free control solution. The agents are trained using historical data generated by an expert system, eliminating the need for direct experimentation with the environment. After deployment, the agents operate autonomously, continuously fine-tuning their policies daily to adapt to evolving process dynamics and reject fast transient disturbances. Experimental validation in an open, industrial-scale $\mathrm{PBR}$ at the University of Almería demonstrated the framework’s capability to maintain stable operation under realistic conditions. The results confirm that model-free MARL control provides a robust and adaptive alternative for complex bioprocess environments.
\end{abstract}


\section{Introduction}
Effective control is paramount in bioprocesses, where living cells act as the fundamental production entities. These cells inherently exhibit complex, autonomous behavior, characterized by internal regulatory mechanisms and heterogeneous distributions within the bioreactor. Such intricate microscale dynamics pose significant control challenges, as they cannot be directly manipulated through standard macroscopic variables. Consequently, maintaining optimal conditions for variables such as nutrient concentration, $\mathrm{pH}$, temperature, and dissolved oxygen ($\mathrm{DO}$) is critical to ensuring cell growth and high productivity, thus necessitating the deployment of advanced control strategies \citep{GuzACC2025}.

Microalgae-based bioprocesses exemplify these control challenges due to their strong dependence on fluctuating environmental and operational conditions. As photosynthetic microorganisms, they convert solar energy and $\mathrm{CO}_2$ into biomass while generating oxygen, with their growth intrinsically linked to the availability of nutrients, light, temperature, $\mathrm{pH}$, and $\mathrm{DO}$ \citep{tarafdar2023environmental}. A key operational task involves precisely managing $\mathrm{CO}_2$ injection, which serves both as the carbon source and as a $\mathrm{pH}$ buffer, alongside air injection, which regulates $\mathrm{DO}$ levels. Thus, maintaining stable $\mathrm{pH}$ and $\mathrm{DO}$ conditions becomes essential, underscoring the necessity for advanced, automated control strategies to achieve efficient and continuous microalgae cultivation \citep{GuzACC2025}.

Regulating $\mathrm{pH}$ in photobioreactors (PBRs) has been extensively studied, leading to the development of numerous robust, adaptive, and model-based control techniques. Conventional approaches, including adaptive Model Predictive Control (MPC) \citep{amaro2023adaptive}, Model Reference Adaptive Control (MRAC) \citep{caparroz2025hybrid}, and more recently, learning-driven control frameworks \citep{pataro2023learning}, have shown reliable performance when the system operates under nominal conditions. Nevertheless, the effectiveness of these control schemes depends heavily on the availability of an accurate process model that must be continuously refined, as well as on prior knowledge of system dynamics and operational constraints. This dependency becomes particularly problematic in microalgal bioprocesses, where ongoing biological and environmental fluctuations cause continuous and complex changes in system behavior. Given this dynamic and uncertain nature, model-free and data-driven control paradigms have emerged as more versatile and promising alternatives, offering enhanced flexibility and adaptability by dispensing with the need for explicitly maintained process models \citep{Haiting2025, gil2025reinforcement}. In contrast, control strategies targeting $\mathrm{DO}$ remain comparatively scarce. The joint regulation of $\mathrm{pH}$ and $\mathrm{DO}$ was initially investigated through a selective control approach by \cite{pawlowski2015selective}; however, this early study did not explicitly address the multivariable interactions inherent to the system. As a result, most current implementations of multivariable control in $\mathrm{PBRs}$ still rely on basic ON/OFF logics \citep{pawlowski2016event} or conventional Proportional-Integral-Derivative (PID) controllers \citep{barcelo2022new}. Such strategies often fail to maintain stable $\mathrm{DO}$ dynamics under the nonlinear and time-varying conditions typical of microalgal photobioreactors. Importantly, the simultaneous regulation of $\mathrm{pH}$ and $\mathrm{DO}$ in open microalgal $\mathrm{PBRs}$ remains a critical and underexplored challenge, representing a highly nonlinear, multivariable, and disturbance-prone control problem.

To address this challenge and harness the potential of model-free control to overcome the difficulties of bioprocess modeling, this study introduces an offline-online Multi-Agent Reinforcement Learning (MARL) framework. The proposed multivariable controller employs Deep Deterministic Policy Gradient (DDPG) agents~\citep{rajasekhar2025exploring}, enabling a fully data-driven and model-free control methodology. Within this approach, the agents are trained using historical data generated by an expert control system composed of PID regulators, thereby eliminating the direct interaction with the physical process. Once trained, the agents are deployed to operate autonomously, collecting process data during daytime operation and undergoing fine-tuning each night. This continual adaptation enables the controller to adjust to evolving system dynamics and to more effectively reject rapid and transient disturbances. Through iterative policy refinement, the offline-trained RL architecture successfully manages the nonlinearities and external perturbations inherent to open $\mathrm{PBRs}$. The proposed approach was  validated in an open, industrial-scale $\mathrm{PBR}$ operated during four days. All experiments were conducted at the UAL-CIESOL research facilities, located at IFAPA center near the University of Almería (UAL). To the best of our knowledge, this represents the first validation of an MARL-based control strategy in an industrial-scale bioprocess.

The remainder of this paper is organized as follows. Section~\ref{Mat} describes the PBR system used as a reference in this study, together with the control methods considered. Section~\ref{Met} describes the proposed methodology. Section~\ref{results} presents the main experimental results obtained in the real PBR system. Finally, Section~\ref{conclusion} summarizes the key findings and concluding remarks of the work.

\section{Material and methods}
\label{Mat}

\subsection{System overview and control problem description}

The experimental facility utilized in this study is a \textit{raceway} PBR system belonging the CIESOL research center and located in the IFAPA research facilities of the Regional Government of Andalusia, near the UAL. This reactor features a surface area of $80~\mathrm{m}^2$ and is composed of two channels, each measuring $40~\mathrm{m}$ in length, $1~\mathrm{m}$ in width, and $0.3~\mathrm{m}$ in depth. Culture mixing and circulation are facilitated by a paddlewheel system, which possesses a $1.2~\mathrm{m}$ diameter and eight blades. Downstream of the paddlewheel, a dedicated sump is utilized for the injection of carbon dioxide ($Q_{\mathrm{CO_2}}$) and air ($Q_{\mathrm{air}}$), enabling the control of $\mathrm{pH}$ and $\mathrm{DO}$, respectively.

The system is comprehensively instrumented, allowing for high-frequency data acquisition (recording every second). This instrumentation captures various essential process variables, including $\mathrm{pH}$, $\mathrm{DO}$, culture temperature, and liquid level, alongside environmental parameters such as solar radiation ($\mathrm{I}$), air temperature, wind speed, and relative humidity. Critical measurements for $\mathrm{pH}$ and $\mathrm{DO}$ are taken at two key locations: immediately following the sump and at the end of the channel, just preceding the paddlewheel. The latter location poses the most significant control challenge, as it is spatially the farthest point from the injection zones for $Q_{\mathrm{CO_2}}$ and $Q_{\mathrm{air}}$, thus serving as the primary focus for the implemented control strategies. A thorough technical description of this system is available in the work by \cite{caparroz2025hybrid}.

\begin{figure}[h]
\centering
  \includegraphics[trim = 130mm 90mm 130mm 55mm, clip, width=0.6\columnwidth]{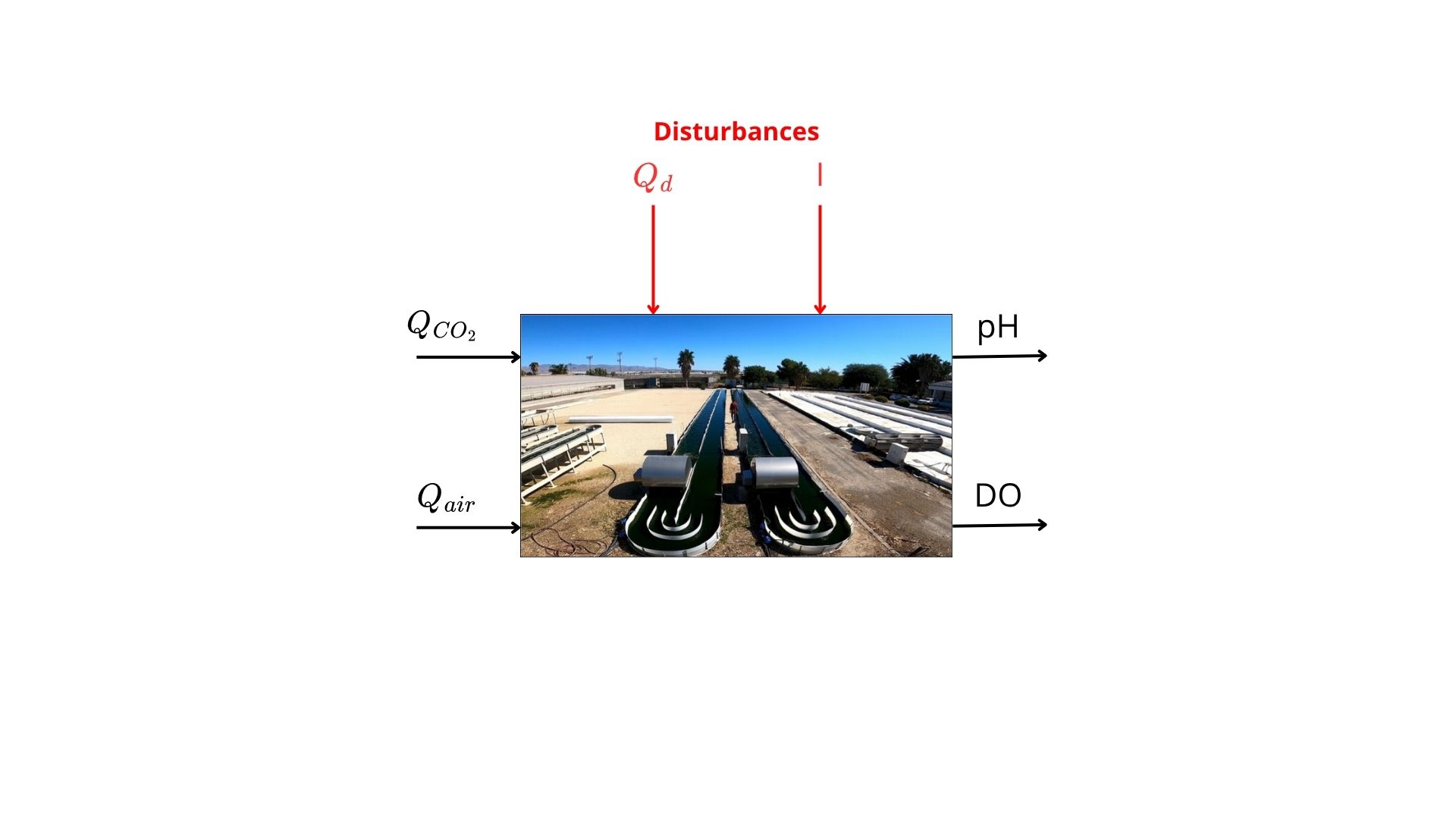}
  \caption{PBR system external representation.}\label{CSTR}
\end{figure}

From an external representation perspective (see Fig.~\ref{CSTR}), the $\mathrm{pH}$ and $\mathrm{DO}$ control problem within the PBR is affected not only by solar irradiance ($\mathrm{I}$) but also by additional factors, notably the dilution flow rate ($Q_{\mathrm{d}}$). This flow is introduced irregularly, either following biomass harvesting or to compensate for liquid losses due to evaporation, and therefore lacks a predefined pattern. Variations in $Q_{\mathrm{d}}$ influence mass transfer and concentration balances, significantly impacting parameters such as $\mathrm{pH}$ and shaping the overall system dynamics and control requirements. The overarching control objective is to maintain optimal $\mathrm{pH}$ and $\mathrm{DO}$ conditions for the cultivated microalgae strain by manipulating the flows of $Q_{\mathrm{CO_2}}$ and $Q_{\mathrm{air}}$. It is essential to recognize the inherent multivariable nature of the process, particularly the observed interactions where high air flow rates ($Q_{\mathrm{air}}$) can highly influence the $\mathrm{pH}$ level.

\subsection{Reinforcement learning background}

In the field of RL, many algorithms rely on the Markov Decision Process (MDP) framework, which assumes that the agent has full observability of the environment's true state $\mathbf{x}_t$ at every time step. However, this assumption is often unrealistic in practical applications. In complex systems, such as bioprocesses, the agent rarely observes the complete state directly. Instead, the agent receives partial observations $\mathbf{o}_t$, which  provide incomplete information regarding the underlying state. To model such environments with limited observability, the Partially Observable MDP (POMDP) framework is employed. A POMDP extends the MDP by explicitly incorporating perceptual uncertainty. The full definition of a POMDP is given by the tuple ($\mathcal{X}$,$\mathcal{O}, \mathcal{U}, P, R, \gamma)$, where: $\mathcal{X}$ is the state space, $\mathcal{O}$ is the observation space, $\mathcal{U}$ represents the action space, $P$ describes the transition dynamics $P(\mathbf{x}_{t+1} \mid \mathbf{x}_t, \mathbf{u}_t)$, $R$ is the reward function $R(\mathbf{x}_t , \mathbf{u}_t)$, and $\gamma \in [0,1]$ is the discount factor. In this framework, the interaction dynamics are sequential: at each discrete time step $t$, the agent receives a partial observation $\mathbf{o}_t$. Based on its policy $\pi$, the agent selects an action $\mathbf{u}_t$. The environment then transitions to a new state $\mathbf{x}_{t+1}$, generates a new observation $\mathbf{o}_{t+1}$, and provides a reward $r_{t}$. 
To develop an RL agent within this POMDP framework, a commonly
adopted approach is the \textit{actor-critic} architecture. The \textit{actor} selects actions $\mathbf{u}_t$ based on the partial observation $\mathbf{o}_t$ via a deterministic policy $\pi(\mathbf{o}_t; \boldsymbol{\theta})$. Concurrently, the \textit{critic} evaluates these actions using an action-value function $Q(\mathbf{o}_t, \mathbf{u}_t; \boldsymbol{\Phi})$. The parameters $\boldsymbol{\theta}$ (actor) and $\boldsymbol{\Phi}$ (critic) are iteratively updated to improve both decision-making and policy valuation.

\subsection{Deep Deterministic Policy Gradient}

Among the most commonly employed \textit{actor-critic} algorithms for 
environments with continuous action spaces is the DDPG 
\citep{rajasekhar2025exploring}. This method utilizes two neural networks: 
an \textit{actor} network that outputs deterministic actions, and a 
\textit{critic} network that evaluates the associated action-value function. 
For offline training, experiences collected over time are stored in a replay 
memory, commonly referred to as an experience \textit{buffer}, which contains 
observations, actions, and rewards sampled at discrete time steps 
\citep{Haiting2025}. During training, random mini-batches of size \(M\) are 
drawn from this buffer to update both the \textit{actor} and \textit{critic} 
networks. The \textit{critic} is updated by minimizing the following loss 
function:
\begin{align}
\label{eq1}
y_i &= r_i + \gamma Q_T(\mathbf{o}_{i+1}, \pi_T(\mathbf{o}_{i+1}; 
       \boldsymbol{\theta}_T); \boldsymbol{\Phi}_T), \\
       \label{eq2}
L(\boldsymbol{\Phi}) &= \frac{1}{M} \sum_{i=1}^{M} \left( y_i - 
       Q(\mathbf{o}_i, \mathbf{u}_i; \boldsymbol{\Phi}) \right)^2.
\end{align}
\noindent Eq.~(\ref{eq1}) defines the temporal-difference target \(y_i\) for 
each sampled transition: \(r_i\) is the immediate reward, and the second 
term provides the discounted estimate of future returns, where the target 
actor \(\pi_T\) selects the next action from \(\mathbf{o}_{i+1}\) and the 
target critic \(Q_T\) evaluates its value, weighted by \(\gamma \in (0,1)\). 
Eq.~(\ref{eq2}) minimizes the mean squared error between these targets and the 
online critic predictions, driving the critic toward estimates consistent 
with the Bellman optimality principle. The target networks \(Q_T\) and 
\(\pi_T\) decouple regression targets from the online parameters, preventing 
the feedback loop that would otherwise destabilize training, and are slowly 
updated toward their online counterparts via a smoothing factor 
\(\tau \in (0,1)\).

The \textit{actor} is updated by ascending the gradient of the expected cumulative reward:
\begin{equation}
   \nabla_{\boldsymbol{\theta}} J \approx \frac{1}{M} \sum_{i=1}^{M} \mathbf{G}_{\mathbf{u}_i} \, \mathbf{G}_{\pi_i}, 
\end{equation}
\noindent where \(\mathbf{G}_{\mathbf{u}_i}\) represents the gradient of the \textit{critic} output with respect to the action produced by the \textit{actor}, and \(\mathbf{G}_{\pi_i}\) is the gradient of the \textit{actor} output with respect to its parameters.

\section{Proposed MARL methodology}
\label{Met}

In this study, a hybrid offline-online MARL framework based on the DDPG algorithm is developed to achieve effective control of both $\mathrm{pH}$ and $\mathrm{DO}$ in microalgae PBR systems. The proposed approach adopts a decentralized control architecture, in which independent DDPG agents are assigned to regulate each process variable, $\mathrm{pH}$ and $\mathrm{DO}$, by manipulating the carbon dioxide injection rate ($Q_{\mathrm{CO_2}}$) and the air flow rate ($Q_{\mathrm{air}}$).

A key stability consideration in decentralized MARL is that each agent treats the other agents' policies as part of the environment, which introduces non-stationarity that can compromise convergence. In the 
proposed framework, this is mitigated through three complementary  mechanisms: (i) offline pre-training on fixed historical data eliminates 
dynamic policy interference during the learning phase; (ii) slow online 
fine-tuning ensures that policy updates remain small and gradual,  preventing abrupt behavioral shifts that could destabilize the coupled 
system; and (iii) soft target network updates further dampen oscillations 
in value estimates. Together, these mechanisms promote stable convergence 
despite the decentralized nature of the training procedure.

\subsection{Offline training}

The proposed methodology for offline training is summarized in 
Algorithm~\ref{Algori1}. In this stage, the offline MARL algorithm 
iteratively updates the \textit{actor} and \textit{critic} networks of 
each agent using historical data generated with the expert PID-based 
controller, without requiring any direct interaction with the physical 
system. Each agent independently learns its control policy by minimizing 
its critic loss and maximizing the expected cumulative reward through 
policy gradient optimization. Although training is decentralized, all 
agents operate within a shared environment and implicitly exchange 
information through process observations, which capture the coupled 
dynamics of the PBR. This shared representation promotes emergent 
coordination among agents, enabling consistent action adaptation and 
balanced regulation of $\mathrm{pH}$ and $\mathrm{DO}$. The 
\textit{critic} networks estimate the expected return for given 
observation-action pairs, whereas the \textit{actor} networks are 
optimized to generate actions that maximize these value estimates. 
Finally, soft target network updates are employed to stabilize training 
and prevent divergence.

\begin{algorithm}[h]
\SetAlgoLined
\KwIn{Set of agents $\mathcal{A} = \{\text{pH}, \text{DO}\}$; 
offline datasets $\mathcal{D}^a = \{(\mathbf{o}_j^a, \mathbf{u}_j^a, r_j^a, \mathbf{o}_{j+1}^a)\}_{j=1}^{N_a}$ for each $a \in \mathcal{A}$; \textit{actor} networks $\pi^a$, \textit{critic} networks $Q^a$, target networks $\pi_T^a$, $Q_T^a$; smoothing factor $\tau$, discount factor $\gamma$, mini-batch size $M$.}
\KwOut{Optimal actor policies $\pi^{pH*}$ and $\pi^{DO*}$.}
\textbf{Initialize:} For each agent $a \in \mathcal{A}$, randomly initialize \textit{actor} and \textit{critic} networks with parameters $\boldsymbol{\theta}^a$, $\boldsymbol{\Phi}^a$; Set target network parameters $\boldsymbol{\theta}_T^a \leftarrow \boldsymbol{\theta}^a$, $\boldsymbol{\Phi}_T^a \leftarrow \boldsymbol{\Phi}^a$.

\For{each training iteration}{
  Sample a mini-batch $\{(\mathbf{o}_i^a, \mathbf{u}_i^a, r_i^a, \mathbf{o}_{i+1}^a)\}_{i=1}^M$ for both $a \in \mathcal{A}$ from $\mathcal{D}^a$\;

  \For{each agent $a \in \mathcal{A}$}{
    Compute target Q-value:\\
    $y_i^a = r_i^a + \gamma\, Q_T^a(\mathbf{o}_{i+1}^a, \pi_T^a(\mathbf{o}_{i+1}^a; \boldsymbol{\theta}_T^a); \boldsymbol{\Phi}_T^a)$
    \\
    Update \textit{critic} by minimizing loss:\\
    $L(\boldsymbol{\Phi}^a) = \frac{1}{M} \sum_{i=1}^{M} \big( y_i^a - Q^a(\mathbf{o}_i^a, \mathbf{u}_i^a; \boldsymbol{\Phi}^a) \big)^2$
    \\
    Update \textit{actors} via policy gradient:\\
  $\nabla_{\boldsymbol{\theta}^{a}} J^{a} \approx \frac{1}{M} \sum_{i=1}^{M} \mathbf{G}_{\mathbf{u}_i^{a}} \, \mathbf{G}_{\pi_i^{a}}$\\
    Soft-update target networks:\\
    $\boldsymbol{\theta}_T^a \leftarrow \tau \boldsymbol{\theta}^a + (1 - \tau)\boldsymbol{\theta}_T^a$,\\ 
    $\boldsymbol{\Phi}_T^a \leftarrow \tau \boldsymbol{\Phi}^a + (1 - \tau)\boldsymbol{\Phi}_T^a$
  }
}
\caption{Offline training of the DDPG-based MARL framework}
\label{Algori1}
\end{algorithm}

\subsection{Online fine-tuning}
To improve adaptability and robustness, the MARL-trained agents undergo an online fine-tuning phase. This stage serves a dual purpose: (i) to adapt the agents’ policies to the time-varying dynamics of the PBR system, and (ii) to enable progressive learning from the influence of external disturbances or process drifts. Through  continuous adaptation, the agents are able to refine their performance beyond that of the baseline PID-type controllers from which their initial knowledge was derived. At the beginning of the fine-tuning process, the replay \textit{buffer} is preloaded with historical data and subsequently updated with new experiences collected during real-time operation. This strategy allows dynamic policy updates based on recent data that capture seasonal variations and other evolving characteristics. The overall fine-tuning procedure is detailed in Algorithm~\ref{algori2}.

\begin{algorithm}[H]
\caption{Online fine-tuning of the DDPG-based MARL framework}
\label{algori2}
\textbf{Input:} Optimal \textit{critic} networks and \textit{actor} policies from the offline training.  

\textbf{Output:} Updated optimal \textit{actor} policies $\pi^{pH*}$ and $\pi^{DO*}$.

\textbf{Initialize:} For each agent $a \in \mathcal{A}$,  initialize \textit{actor} and \textit{critic} networks with the optimal parameters obtained from the offline training $\boldsymbol{\theta}^a \leftarrow \boldsymbol{\theta}^{*, a} $, $\boldsymbol{\Phi}^a \leftarrow \boldsymbol{\Phi}^{*, a}$. Initialize each agent's experience \textit{buffer} using the corresponding historical dataset \(\mathcal{D}^a\).

\vspace{2pt}
\textbf{For fine-tuning of the DDPG agents do}
\begin{enumerate}
\setlength{\itemsep}{0pt}
    \item During PBR operation, at each sampling time \(t\), collect new experiences for each agent: $\mathcal{D}^{a}_t = (\mathbf{o}_t, \mathbf{u}_t, r_t, \mathbf{o}_{t+1}) , \quad \forall a \in \mathcal{A}$. The data set of new experiences will be given by: 
    \[
        \mathcal{D}^{New, a} = \{(\mathbf{o}_j, \mathbf{u}_j, r_j, \mathbf{o}_{j+1})\}_{j=1}^{N^{New, a}}, \quad \forall a \in \mathcal{A}.
    \]
    \item Update each agent's \textit{buffer} dynamically, keeping the most recent experiences.
    \item Fine-tune all agents using the same training procedure as in Algorithm~\ref{Algori1}, applied concurrently to each agent \(a \in \mathcal{A}\). The fine-tuning may use a reduced number of epochs to avoid overfitting or destabilization.
\end{enumerate}
\end{algorithm}

\section{Results}
\label{results}

This section reports the outcomes derived from applying the proposed hybrid MARL-based control methodology to the real PBR system. 

\subsection{Collection of historical experiences}

\begin{figure}[h]
\centering
  \includegraphics[trim = 40mm 50mm 100mm 30mm, clip, width=0.8\columnwidth]{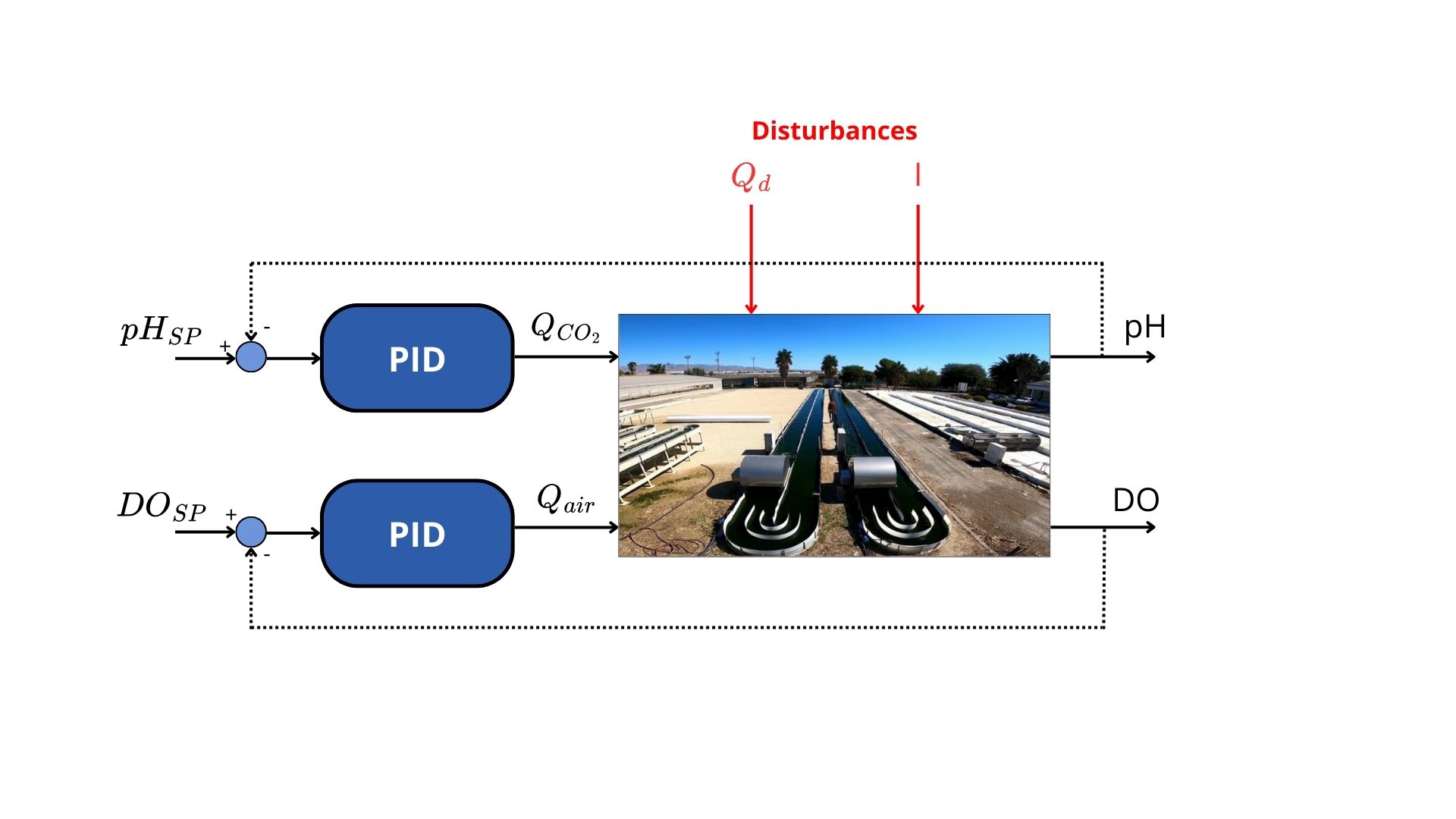}
  \caption{PID-based control scheme for the PBR system.}\label{figPID}
\end{figure}

The DDPG agents were trained using historical datasets obtained from the PBR operating under PID-based control, see Fig.~\ref{figPID}. This expert system included two independent PID controllers dedicated to regulating the $\mathrm{pH}$ and $\mathrm{DO}$ variables, respectively, by manipulating the carbon dioxide injection rate ($Q_{\mathrm{CO_2}}$) and the air flow rate ($Q_{\mathrm{air}}$). It is important to note that this scheme did not incorporate auxiliary components typically found in multivariable control systems, such as decouplers or feedforward controllers for interaction and disturbance rejection, respectively. Both controllers were implemented in their ideal form, excluding the derivative term. The tuning parameters were set to \(K_p = -32~[\mathrm{L/min}]\) and \(T_i = 1200~[\mathrm{s}]\) for the $\mathrm{pH}$ loop, and \(K_p = -2.81~[\mathrm{L/(min\%)}]\) and \(T_i = 600~[\mathrm{s]}\) for the $\mathrm{DO}$ loop. The sampling interval was fixed at 10~s.

\begin{figure}[ht]
\centering
\includegraphics[
    width=16 cm,
    trim={0cm 0cm 0cm 0cm},  
    clip  
]{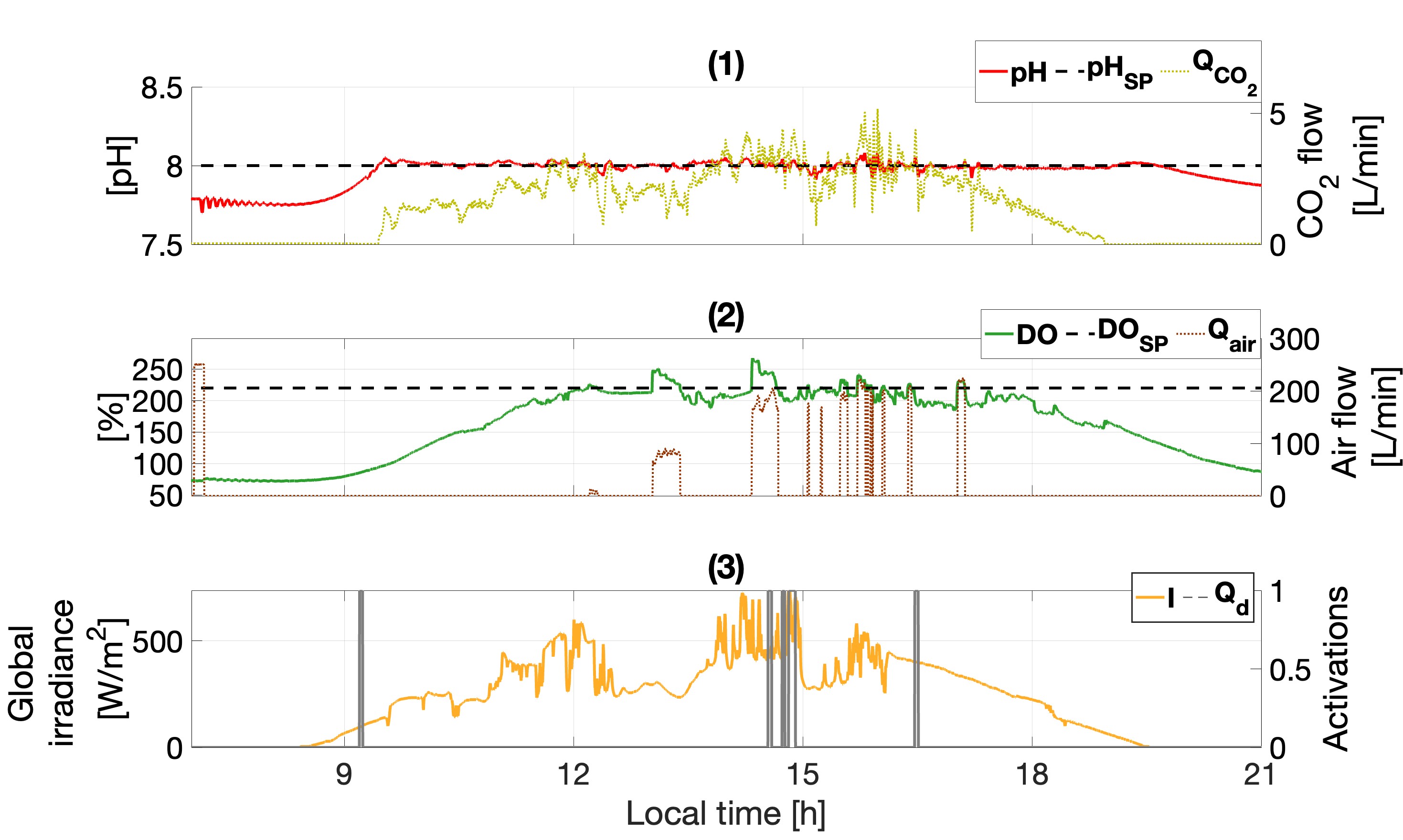}
\caption{PBR operation using the PID controllers.}
\label{fig_real_pid}
\end{figure}

The data were collected over a three-day period (October 11-13, 2025), and one representative day is shown in the Fig.~\ref{fig_real_pid}. During operation, a set of procedural rules was applied. At the beginning of each day, before solar irradiance increased (around 7:00~a.m., see Fig.~\ref{fig_real_pid}-(2)), an air injection was performed to resuspend the biomass accumulated overnight in the sump. Once irradiance exceeded 100~W/m$^2$, the automatic control mode was activated. Under this mode, the PID controllers maintained $\mathrm{pH}$ and $\mathrm{DO}$ around 8 and 220~\%, respectively, which are the optimal operating points for the selected microalgal strain (\textit{Scenedesmus almeriensis}).  During the three-day period, the maximum deviation from the reference was 0.12 [-] for $\mathrm{pH}$ and 47 [\%] for $\mathrm{DO}$.

\subsection{MARL computational implementation}

The historical dataset acquired during the PID-controlled operation provided the basis for training the DDPG agents. This dataset inherently contained the system’s complex, coupled, and disturbance-driven dynamics. In particular, the recorded effects of irradiance, dilution flow, and the strong interaction between control loops were key considerations in defining each agent’s observation space.

\begin{figure}[H]
\centering
  \includegraphics[trim = 35mm 10mm 30mm 5mm, clip, width=0.9\columnwidth]{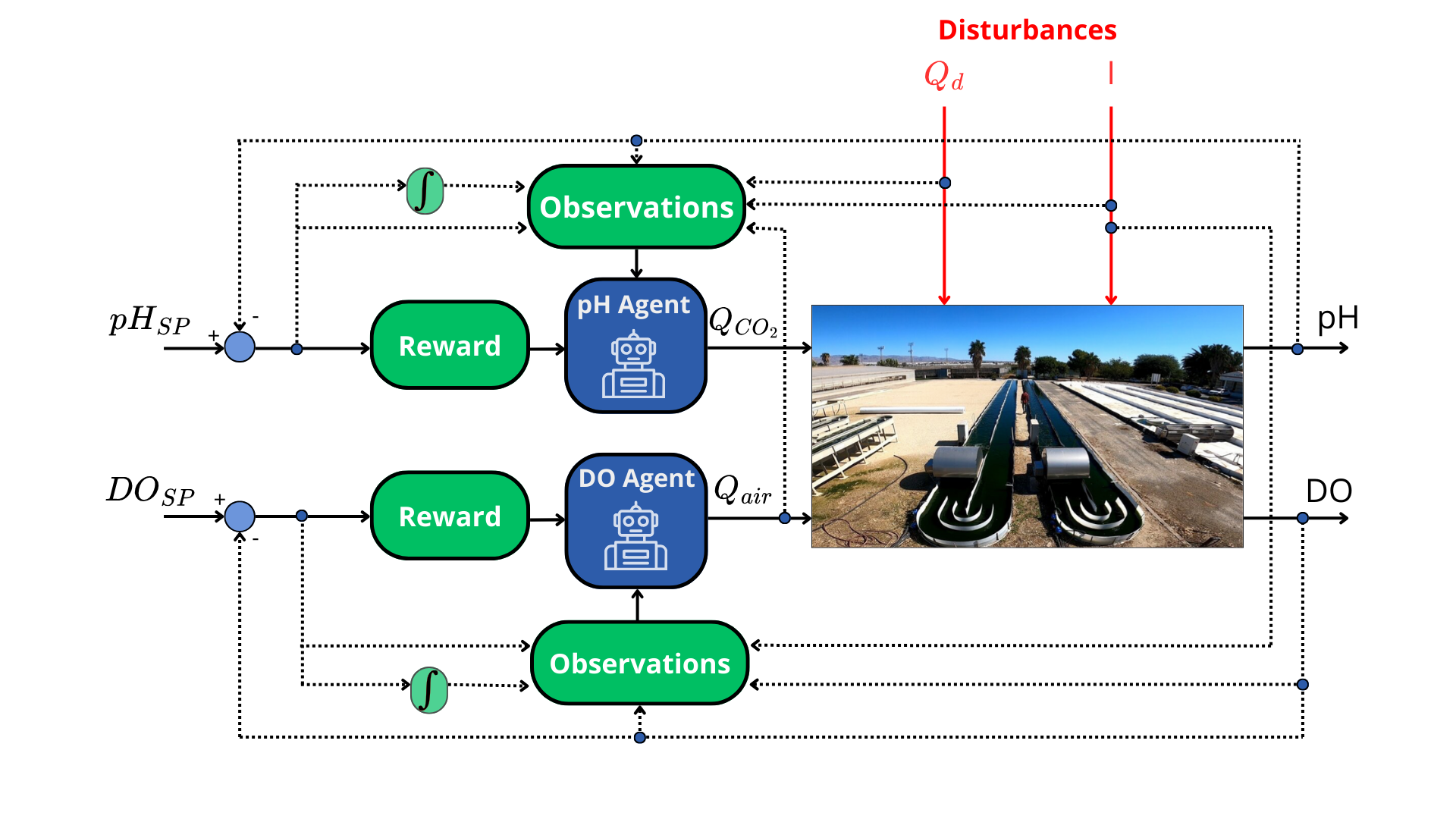}
  \caption{MARL-based control scheme for the PBR system.}\label{figMARL}
\end{figure}

For every agent, classical control-engineering features were incorporated into the observation vector, including the control error, defined as the difference between the setpoint and the measured variable, and the integral of this error. These elements were included to supply the agents with the same type of fundamental information used by the expert PID system that generated the training data. In addition, the observation space for the $\mathrm{pH}$ agent was expanded to include measurements of $\mathrm{pH}$, irradiance, dilution flow rate, and air injection events. Conversely, the agent responsible for $\mathrm{DO}$ regulation received $\mathrm{DO}$ and irradiance as input variables. This configuration was derived from the interactions identified during the PID experiments and from previous experience with other control strategies. The goal was for the agents not only to replicate the expert system’s behavior but also to enhance it by implicitly developing feedforward compensation for dominant disturbances and decoupling actions for the observed inter-loop interactions. Based on these considerations, the overall configuration of the MARL control framework is illustrated in Fig.~\ref{figMARL}. In this scheme, the reward function for each agent was defined as:
\begin{equation}
r_t^a = -\log{\left( (e_t^a)^2 + \epsilon \right)},
\end{equation}
where $e_t^a = Y_{\mathrm{SP}}^a - Y_t^a$ denotes the control error of agent $a$ (with $a \in \{\mathrm{pH}, \mathrm{DO}\}$) at sampling instant $t$, calculated as the difference between the desired setpoint $Y_{\mathrm{SP}}^a$ and the measured output $Y_t^a$. This logarithmic reward formulation was adopted to avoid numerical instabilities during gradient computation, as discussed in \cite{gil2025reinforcement}.

Both agents were implemented in MATLAB~\cite{MATLAB2019}. The \textit{actor} network of each agent comprised eight layers of 256 neurons each with different activation functions, including \textit{ReLU} and \textit{Tanh}, whereas the \textit{critic} network consisted of nine layers with comparable complexity. In terms of DDPG hyperparameter settings, both agents utilized the Adam optimizer~\citep{kingma2014adam}, with learning rates of $10^{-3}$ for the \textit{critic} and $10^{-4}$ for the \textit{actor} networks. The discount factor ($\gamma$) was set to 0.9, the target update coefficient ($\tau$) to 0.01, and the mini-batch size ($M$) to 64.

\begin{figure}[!h]
\centering
  \includegraphics[width=0.7\linewidth, clip, trim=12cm 1cm 12cm 1.5cm]{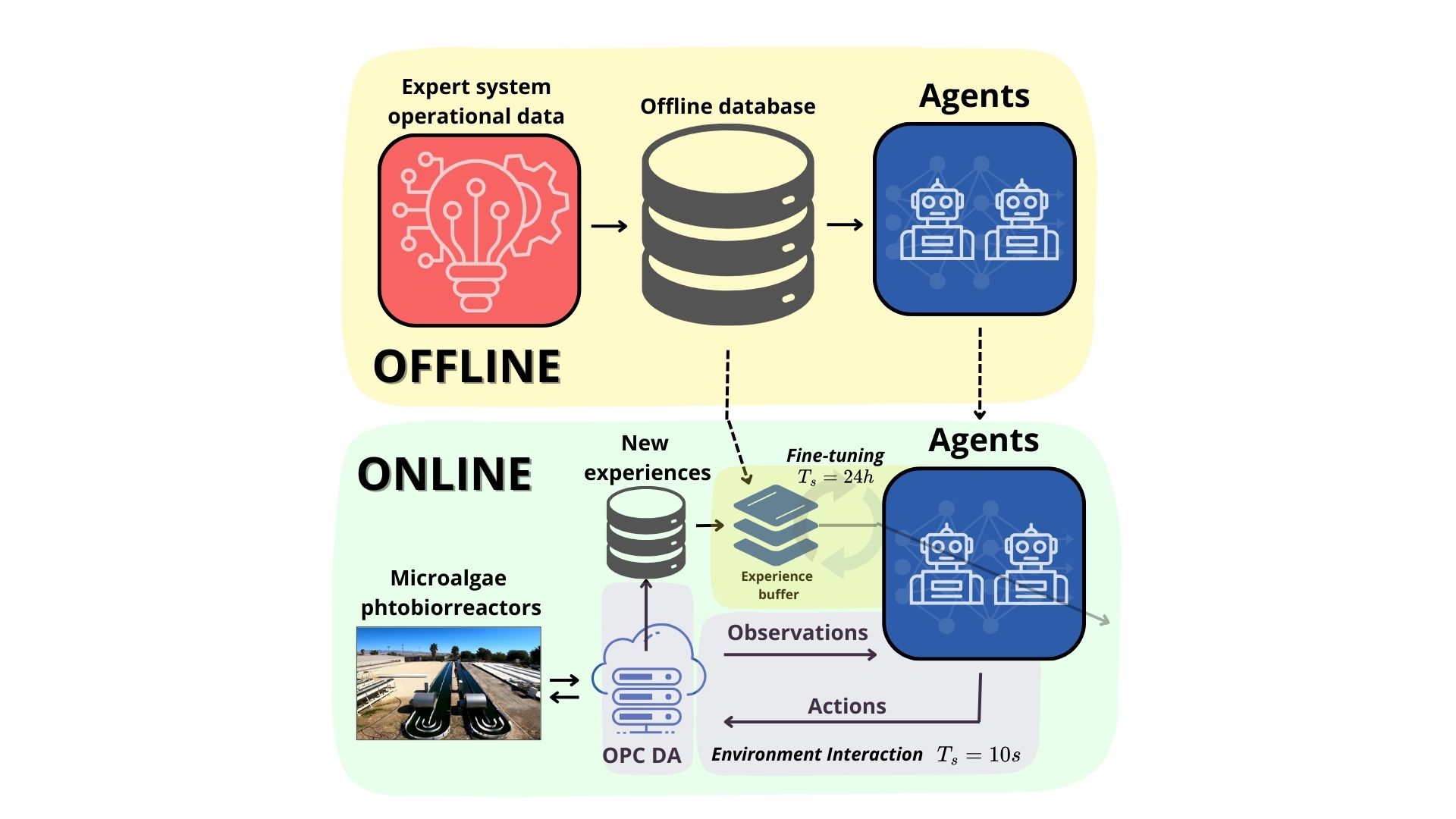}\\
  \caption{Hybrid MARL framework implementation.}\label{figHy}
\end{figure}

The entire implementation for transitioning from the offline to the online phase was carried out following the scheme shown in Fig.~\ref{figHy}. During the offline training stage, the sampling interval was fixed at 10~s to maintain consistency with the PID-based control experiments, and the training process was executed for up to 3000 epochs. In the online phase, communication with the real PBR system was established through an OPC server, and the fine-tuning of the agents was performed at the end of each operating day using 50 epochs to avoid destabilization.

\subsection{Results of the MARL approach in the real PBR system}

\begin{figure*}[ht]
\centering
\includegraphics[
    width=16.2cm,
    trim={0cm 0cm 0cm 0cm},  
    clip  
]{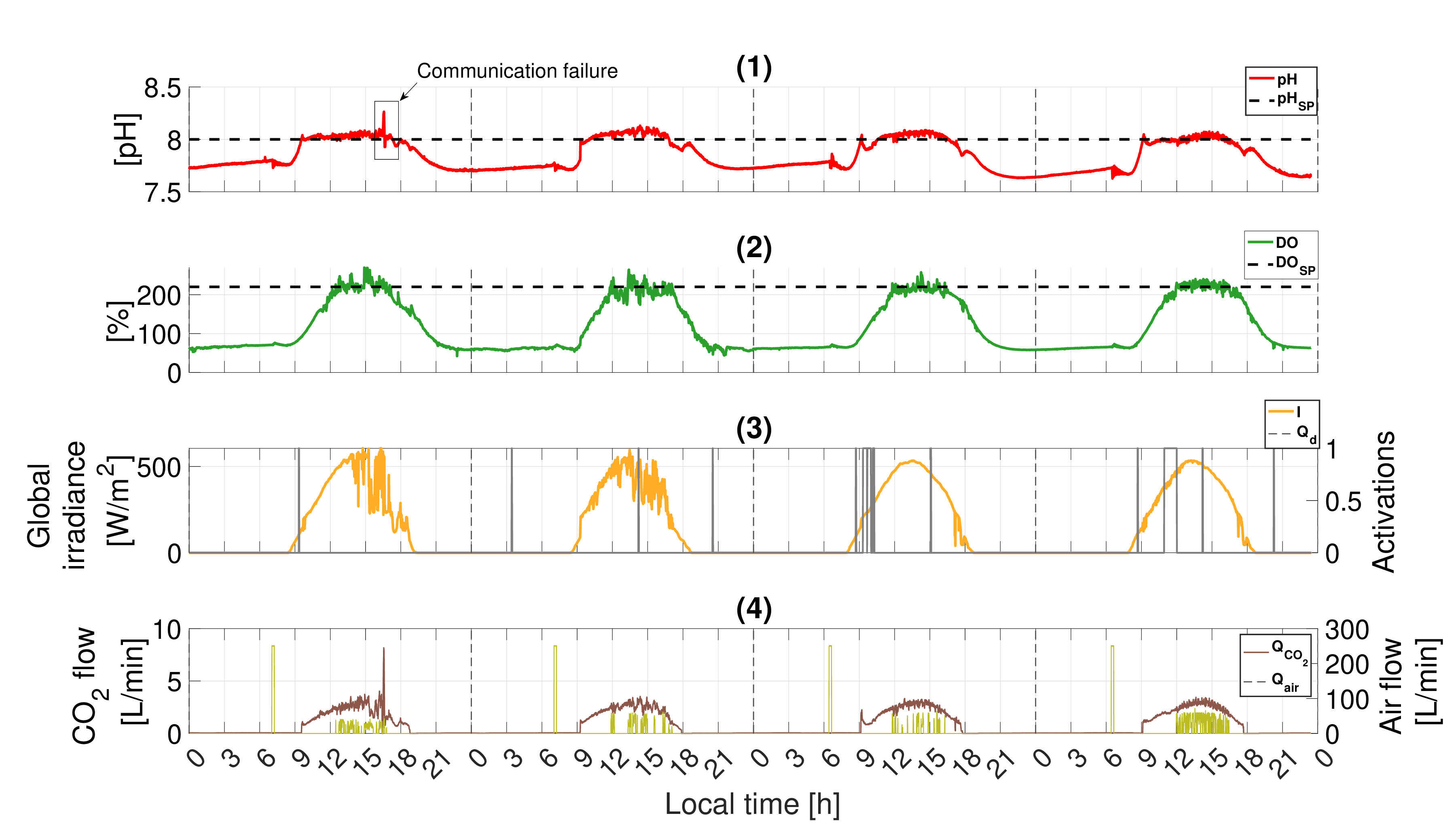}
\caption{Operation of the PBR system using the proposed MARL controller. }
\label{fig_resultados_finales}
\end{figure*}

The experiments with the proposed MARL controller were conducted between 19 and 22 October 2025, and the corresponding results are shown in Fig.~\ref{fig_resultados_finales}. Fine-tuning was performed at the end of each day, as indicated by the vertical dashed lines in the figure. The first day of operation corresponded to a weekend, during which no daily operation activities, such as reactor harvesting, were carried out, although irradiance fluctuations and a brief communication failure occurred toward the end of the day. Despite these disturbances, both $\mathrm{pH}$ and DO control remained stable and within acceptable limits around the reference (see Fig.~\ref{fig_resultados_finales}-(1) and (2)). 

On the second day, even though it was a working day, no harvesting operations were performed, resulting in fewer dilution flow injections (see Fig.~\ref{fig_resultados_finales}-(3)). Passing clouds were again observed (see Fig.~\ref{fig_resultados_finales}-(3)), and both temperature and irradiance reached lower values than those recorded during the training phase, resulting in a change in the dynamic of the system. Consequently, a slight deviation of the $\mathrm{pH}$ from its reference value was observed around midday (see Fig.~\ref{fig_resultados_finales}-(1)). In the following days, after applying the daily fine-tuning, this deviation was no longer present. This effect is most evident on the third day, when harvesting was carried out at the beginning of the day, introducing a dilution flow (see Fig.~\ref{fig_resultados_finales}-(3)) that caused a temporary drop in $\mathrm{pH}$ (see Fig.~\ref{fig_resultados_finales}-(1)). However, on the final day, under similar conditions, the fine-tuned controller effectively compensated for this disturbance, achieving accurate regulation of both $\mathrm{pH}$ and $\mathrm{CO_2}$ concentrations (see Fig.~\ref{fig_resultados_finales}-(1) and (2)).

Considering all these observations, the largest deviation from the reference recorded during the four days of operation was 0.12~[-] for $\mathrm{pH}$ and 44~[\%] for $\mathrm{DO}$. These values are comparable to those achieved with the PID-based control, demonstrating that the proposed MARL controller can maintain equivalent performance while providing superior adaptability under varying operational conditions.

\section{Conclusion}
\label{conclusion}

This work presented a hybrid offline-online MARL framework for the simultaneous control of $\mathrm{pH}$ and $\mathrm{DO}$ in open microalgae PBR systems. The proposed approach employed two decentralized DDPG agents, each responsible for one process variable, trained initially with historical data from an expert PID-based control system and subsequently fine-tuned online using real process data. Experimental validation over four consecutive days in an industrial-scale PBR demonstrated that the MARL-based controller achieved stable regulation of both $\mathrm{pH}$ and $\mathrm{DO}$ under realistic operational conditions, such as fluctuations in irradiance or dilution flow injections. The daily fine-tuning stage effectively improved adaptability, enabling the agents to correct deviations and compensate for variable coupling effects. These results confirm that data-driven, model-free MARL control constitutes a robust and adaptive alternative for complex bioprocess environments. 

Future work will focus on evaluating the proposed methodology over a full year to assess long-term adaptability under seasonal variability. Additionally, further developments will explore architectures with a shared \textit{critic} to mitigate potential non-stationarity issues and extend the framework toward cooperative multi-agent configurations.

\section*{Acknowledgment}
This project is part of the R\&D\&I project PID2023-150739OB-I00, funded by MCIN/ AEI/10.13039/501100011033/ and ``FEDER A way to make Europe", and also by the European Union (Grant agreement ID: 101060991, REALM.

\bibliographystyle{unsrtnat}
\bibliography{references.bib}  






\end{document}